\documentclass[twocolumn]{aa}
\usepackage{graphicx}
\begin{document}

\authorrunning{Nagao et al.}
\titlerunning{Ly$\alpha$ emitters with a large equivalent width}

\title{High-redshift Ly$\alpha$ emitters with a large equivalent width}
\subtitle{Properties of $i^\prime$-dropout galaxies with
          an NB921-band depression \\ in the Subaru Deep Field}

\author{
          Tohru Nagao            \inst{1, 2},
          Takashi Murayama       \inst{3},
          Roberto Maiolino       \inst{1, 4},
          Alessandro Marconi     \inst{1, 5},
          Nobunari Kashikawa     \inst{2}, \\
          Masaru Ajiki           \inst{3}, 
          Takashi Hattori        \inst{6}, 
          Chun Ly                \inst{7},
          Matthew A. Malkan      \inst{7},
          Kentaro Motohara       \inst{8}, \\
          Kouji Ohta             \inst{9},
          Shunji S. Sasaki       \inst{3, 10},
          Yasuhiro Shioya        \inst{10}, \and
          Yoshiaki Taniguchi     \inst{10}
}

\offprints{T. Nagao}

\institute{
           INAF -- Osservatorio Astrofisico di Arcetri,
           Largo Enrico Fermi 5, 50125 Firenze, Italy\\
           \email{tohru@arcetri.astro.it}
           \and
           National Astronomical Observatory of Japan,
           2-21-1 Osawa, Mitaka, Tokyo 151-8588, Japan
           \and
           Astronomical Institute, Graduate School of Science,
           Tohoku University, Aramaki, Aoba, Sendai 980-8578, Japan
           \and
           INAF -- Osservatorio Astrofisico di Roma,
           Via di Frascati 33, 00040 Monte Porzio Catone, Italy
           \and
           Dipartimento di Astronomia e Scienza dello Spazio,
           Universit\`{a} di Firenze, Largo E. Fermi 2, 50125 Firenze, Italy
           \and
           Subaru Telescope, National Astronomical Observatory of Japan, 
           650 North A'ohoku Place, Hilo, HI 96720, USA
           \and
           Department of Astronomy, University of California at Los Angeles, 
           P. O. Box 951547, Los Angeles, CA 90095-1547, USA
           \and
           Institute of Astronomy, Graduate School of Science, 
           University of Tokyo, 2-21-1 Osawa, Mitaka, Tokyo 181-0015, Japan
           \and
           Department of Astronomy, Graduate School of Science, 
           Kyoto University, Kitashirakawa, Sakyo, Kyoto 606-8502, Japan
           \and
           Department of Physics, Graduate School of Science and Engineering,
           Ehime University, 2-5 Bunkyo-cho, Matsuyama 790-8577, Japan
}

\date{Received ;  accepted }

\abstract{
  We report new follow-up spectroscopy of $i^\prime$-dropout 
  galaxies with an NB921-band depression found in the Subaru 
  Deep Field. The NB921-depressed $i^\prime$-dropout selection 
  method is expected to select galaxies with large equivalent 
  width Ly$\alpha$ emission over a wide redshift range,
  $6.0 \la z \la 6.5$. 
  Two of four observed targets
  show a strong emission line with a clear asymmetric profile,
  identified as Ly$\alpha$ emitters at $z$ = 6.11 and 6.00.
  Their rest-frame equivalent widths are 153$\mbox{\rm \AA}$ and 
  114$\mbox{\rm \AA}$, which are lower limits on the intrinsic 
  equivalent widths. 
  Through our spectroscopic observations (including 
  previous ones) of NB921-depressed $i^\prime$-dropout galaxies, 
  we identified 5 galaxies in total with a rest-frame equivalent 
  width larger than 100$\mbox{\rm \AA}$ at $6.0 \la z \la 6.5$ out of
  8 photometric candidates, which suggests that the 
  NB921-depressed $i^\prime$-dropout selection method is 
  possibly an efficient way to search for Ly$\alpha$ emitters 
  with a large Ly$\alpha$ equivalent width, in a wider redshift 
  range than usual narrow-band excess techniques.
  By combining these findings with our previous observational 
  results, we infer that the fraction of broad-band selected 
  galaxies having a rest-frame equivalent width larger than 
  100$\mbox{\rm \AA}$ is significantly higher at $z \sim 6$ 
  (the cosmic age of $\sim$1 Gyr) than that at $z \sim 3$
  ($\sim$2 Gyr), 
  being consistent with the idea
  that the typical stellar population of galaxies is 
  significantly younger at $z \sim 6$ than that at $z \sim 3$. 
  The NB921-depressed $i^\prime$-dropout galaxies may be 
  interesting candidates for hosts of massive, zero-metallicity 
  Population III stars.
}

\maketitle

\section{Introduction}

\begin{table*}
\centering
\caption{Photometric properties of NB921-depressed 
         $i^\prime$-dropout galaxies in the Subaru Deep Field}
\label{table:01}
\begin{tabular}{c l c c c c c c}
\hline\hline
\noalign{\smallskip}
No.                      &
Name                     & 
Redshift                 &
$i^\prime$$^\mathrm{a}$  &
$z^\prime$$^\mathrm{a}$  &
$NB921$$^\mathrm{a}$     &
$z^\prime \! - \! NB921$ &
Ref.$^\mathrm{b}$        \\
\noalign{\smallskip}
\hline 
\noalign{\smallskip}
1 & SDF J132345.6+271701 & 6.11 &$>$27.9& 25.24 & 26.37 &  --1.13 & 1 \\
2 & SDF J132422.0+271742 &---   &$>$27.9& 25.96 &$>$27.0& $<-1.0$ & 1 \\
3 & SDF J132426.5+271600 & 6.03 & 27.43 & 25.36 & 25.92 &  --0.56 & 2 \\
4 & SDF J132440.6+273607 & 6.33 &$>$27.9& 25.66 & 26.20 &  --0.54 & 3 \\
5 & SDF J132442.5+272423 & 6.04 & 27.69 & 25.74 & 26.71 &  --0.97 & 2 \\
6 & SDF J132519.4+271829 & 6.00 &$>$27.9& 25.42 & 26.38 &  --0.96 & 1 \\
7 & SDF J132521.6+274229 &---   &$>$27.9& 25.19 & 26.49 &  --1.30 &---\\
8 & SDF J132526.1+271902 &---   & 27.50 & 24.73 & 24.98 &  --0.25 & 1 \\
\noalign{\smallskip}
\hline
\end{tabular}
\begin{list}{}{}
\item[$^{\mathrm{a}}$]
  AB magnitude with a 2$^{\prime\prime}$ aperture photometry.
  Lower limits are 2$\sigma$ limiting magnitudes.
\item[$^{\mathrm{b}}$]
  References. ---
    1: This work,
    2: Nagao et al. (2005a),
    3: Nagao et al. (2004).
\end{list}
\end{table*}

\begin{figure}
\centering
\includegraphics[width=8.7cm]{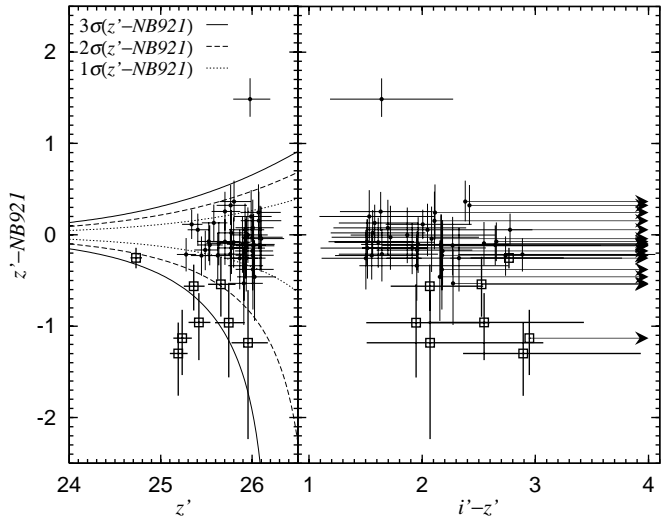}
\caption{
   Color-magnitude diagram of $z^\prime - NB921$ vs. $z^\prime$
   ($left$) and $z^\prime - NB921$ vs. $i^\prime - z^\prime$
   color-color diagram ($right$). The 48 $i^\prime$-dropout 
   objects detected in the SDF are plotted. 
   The NB921-depressed $i^\prime$-dropout objects are shown by 
   open squares, and the other $i^\prime$-dropout objects are 
   shown by filled circles. Error bars denote 1$\sigma$ 
   uncertainty in the observed color or magnitude. In the 
   left panel, 1, 2, and 3 $\sigma$ uncertainties of the 
   $z^\prime - NB921$ color are shown by the dotted, dashed, 
   and solid lines, respectively.
}
\label{fig01}
\end{figure}

Observational searches for high-$z$ galaxies have achieved
important progress in this decade. Thanks to large
telescopes such as Subaru, VLTs and Keck, a few dozens of 
galaxies at $6<z<7$ have been spectroscopically confirmed 
so far (Hu et al. 2002; Kodaira et al. 2003; Cuby et al. 2003; 
Rhoads et al. 2004; Kurk et al. 2004; Nagao et al. 2004, 
2005a; Taniguchi et al. 2005; Stern et al. 2005; 
Kashikawa et al. 2006). Statistical properties of these 
high-$z$ galaxies, such as their luminosity function and 
their correlation function provide information on galaxy 
evolution, the cosmic star-formation rate, and the 
re-ionization history of the universe. In addition to 
statistical properties, properties of individual galaxies 
at high redshift also shed light on galaxy evolution. In 
particular, the equivalent width ($EW$) of the Ly$\alpha$ 
emission provides clues on the stellar population of 
high-$z$ galaxies, for which it is generally difficult to 
investigate stellar spectral features due to the limited 
observational sensitivity of currently available instruments.
Malhotra \& Rhoads (2002) presented their theoretical
estimates that large Ly$\alpha$ equivalent widths
[$EW_0 \ga 150 \mbox{\rm \AA}$ for a Salpeter initial-mass function
(IMF) and $EW_0 \ga 240 \mbox{\rm \AA}$ for a flatter IMF]
suggest a very young ($< 10^7$ years) stellar population 
(e.g., Tumlinson et al. 2003). More interestingly, it has 
been theoretically predicted that galaxies hosting 
zero-metallicity stars (or Population III stars; hereafter 
PopIII) should show huge Ly$\alpha$ equivalent widths which 
could reach up to $EW_0 \sim 1000 \mbox{\rm \AA}$ (e.g., 
Schaerer 2002, 2003; Tumlinson et al. 2003; 
Scannapieco et al. 2003). Therefore, the frequency 
distribution of the Ly$\alpha$ equivalent width in high-$z$ 
galaxies and its evolution with redshift are important in 
understanding the early stages of galaxy evolution 
(e.g., Shimasaku et al. 2006; Ando et al. 2006).

However, the observational study of the Ly$\alpha$ 
equivalent width in galaxies at high redshift is not 
straightforward. Although Malhotra \& Rhoads (2002) 
reported that more than half of their sample of 150 
Ly$\alpha$ emitters (LAEs) have Ly$\alpha$ equivalent 
width larger than 240$\mbox{\rm \AA}$, their analysis is based 
only on photometric data. Especially for LAEs at $z\sim6.5$
(which corresponds to the window between airglow emission 
at $\lambda \sim 9200\mbox{\rm \AA}$), the $z^\prime$-band 
magnitude is contaminated by the Ly$\alpha$ emission, and
thus an unreliable measure of continuum flux density.
Even for the sample with spectroscopic data, the continuum 
emission of LAEs selected by using narrow-band magnitude 
is generally too faint to be detected in their spectra.

We are exploiting a new selection method to identify 
LAEs with a large Ly$\alpha$ equivalent width, that is,
$i^\prime$ dropout with a ``depression'' in the 
narrow-band filter NB921 (see \S2) (Nagao et al. 2004, 
2005a). If galaxies 
are at $6.0 < z < 6.5$, their redshifted Ly$\alpha$ 
emission is expected at 
$8500\mbox{\rm \AA} \la \lambda_{\rm obs} \la 9100\mbox{\rm \AA}$.
In this case, the narrow-band magnitude at 
$\lambda \sim 9200\mbox{\rm \AA}$ has no contamination from
Ly$\alpha$ emission and thus traces only the continuum,
while the $z^\prime$-band magnitude is enhanced by the 
contribution of the strong Ly$\alpha$ emission and thus 
the narrow-band magnitude is depressed with respect to 
the $z^\prime$-band magnitude. Therefore this selection 
method finds galaxies with a large Ly$\alpha$ equivalent 
width in a wide redshift range, $6.0 < z < 6.5$. Note
that this selection method does not select Galactic 
late-type stars, unlike usual $i^\prime$-dropout selection
(see Nagao et al. 2005a for more details).

In this paper, we report new spectroscopy of a sample of 
NB921-depressed $i^\prime$-dropout galaxies. Throughout 
this paper, we adopt a cosmology with 
($\Omega_{\rm tot}$, $\Omega_{\rm M}$, $\Omega_\Lambda$) 
= (1.0, 0.3, 0.7) and $H_0$ = 70 km s$^{-1}$ Mpc$^{-1}$.

\section{Target selection and observation}

\begin{figure}
\centering
\includegraphics[width=7.5cm]{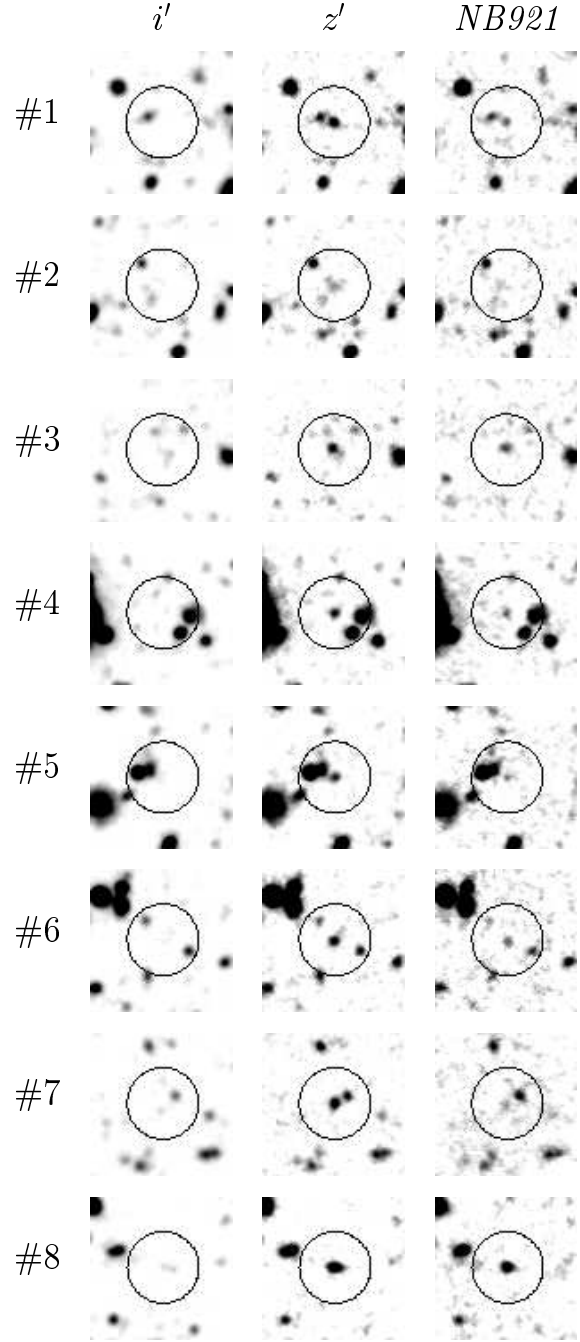}
\caption{
   Thumbnail images of NB921-depressed $i^\prime$-dropout 
   galaxies. The square regions around each object in the 
   $i^\prime$, $z^\prime$, and NB921 images are shown from 
   left to right. The IDs given at the left side of the 
   panels correspond to those in Table 1. The size of 
   panels and the radius of circles are 16 arcsec and 8 
   arcsec, respectively.
}
\label{fig02}
\end{figure}

We selected our $i^\prime$-dropout galaxy sample from 
the public catalog of the Subaru Deep Field (SDF) imaging
survey (Kashikawa et al. 2004), which contains broad-band
($B$, $V$, $R_{\rm C}$, $i^\prime$, and $z^\prime$) and 
narrow-band photometric data [$NB816$ and $NB921$; the 
central wavelengths and the half-widths of the 
transmittance are (8150$\mbox{\rm \AA}$, 120$\mbox{\rm \AA}$) and
(9196$\mbox{\rm \AA}$, 132$\mbox{\rm \AA}$), respectively]. The 
adopted criteria to select the $i^\prime$-dropout galaxy 
sample are:
\begin{itemize}
   \item $z^\prime < 26.1$ (i.e., above 5$\sigma$ 
         background error),
   \item $i^\prime - z^\prime > 1.5$,
   \item $B > 28.5$ (below 3$\sigma$ background error), 
and
   \item $R_{\rm C} > 27.8$ (below 3$\sigma$ background error).
\end{itemize}
Among the 48 selected $i^\prime$-dropout galaxies in SDF,
we identified as a ``NB921-depressed $i^\prime$-dropout 
galaxy'' an object whose $z^\prime - NB921$ is less than
$-2\sigma$ of the sky noise. Here ``sky noise'' refers 
to the error on the $z^\prime - NB921$ color due to sky 
noise in the $z^\prime$ and $NB921$ filters.
The $z^\prime - NB921$ versus $z^\prime$ 
color-magnitude diagram and the $z^\prime - NB921$ 
versus $i^\prime - z^\prime$ color-color diagram are 
shown in Figure 1. Eight $i^\prime$-dropout galaxies 
satisfy this criterion and thus are classified as 
NB921-depressed $i^\prime$-dropout galaxies. Note that 
the significance level of the NB921 depression of one of 
the eight NB921-depressed $i^\prime$-dropout galaxies 
(SDF J132440.6+273607; Nagao et al. 2004) is slightly 
less than 2$\sigma$. However we included this object in 
our NB921-depressed $i^\prime$-dropout galaxy sample.
The object names and the photometric properties of this 
sample are summarized in Table 1, and the thumbnails of 
these objects are shown in Figure 2.

\begin{table*}
\centering
\caption{Spectroscopic properties of NB921-depressed 
         $i^\prime$-dropout galaxies in the Subaru Deep Field}
\label{table:02}
\begin{tabular}{c c c c c c c}
\hline\hline
\noalign{\smallskip}
No.                             &
Redshift                        &
$F$(Ly$\alpha$)$^\mathrm{a}$    &
FWHM$^\mathrm{b}$               &
$EW_0$(Ly$\alpha$)$^\mathrm{c}$ &
$EW_0^{\rm min}$(Ly$\alpha$)$^\mathrm{d}$ &
Ref.$^\mathrm{e}$              \\
                                &
                                &
(10$^{-17}$ ergs/s/cm$^2$)      &
(km s$^{-1}$)                   &
($\mbox{\rm \AA}$)                   &
($\mbox{\rm \AA}$)                   &
                               \\
\noalign{\smallskip}
\hline 
\noalign{\smallskip}
1 & 6.11 & 4.5$\pm$0.1 & 330            & 153 &  72 & 1 \\
3 & 6.03 & 3.6$\pm$0.3 & 410            &  94 &  83 & 2 \\
4 & 6.33 & 4.0$\pm$0.1 &---$^\mathrm{f}$& 130 & 133 & 3 \\
5 & 6.04 & 4.5$\pm$0.3 & 220            & 236 & 146 & 2 \\
6 & 6.00 & 3.4$\pm$0.2 & 220            & 114 &  90 & 1 \\
\noalign{\smallskip}
\hline
\end{tabular}
\begin{list}{}{}
\item[$^{\mathrm{a}}$]
  No correction for the slit loss.
\item[$^{\mathrm{b}}$]
  Corrected for the instrumental broadening 
  (FWHM = 6.2$\mbox{\rm \AA}$).
\item[$^{\mathrm{c}}$]
  Not corrected for the absorption effects due to the 
  Ly$\alpha$-forest, and thus lower limits on the
  intrinsic equivalent width.
\item[$^{\mathrm{d}}$]
  Minimum $EW_0$(Ly$\alpha$) that can be selected by 
  the NB921-depressed $i^\prime$-dropout method for galaxies 
  with the redshift and the $z^\prime$-band magnitude of 
  each NB921-depressed $i^\prime$-dropout galaxy, estimated by
  assuming a flat UV slope (see \S\S4.1).
\item[$^{\mathrm{e}}$]
  References. ---
    1: This work,
    2: Nagao et al. (2005a),
    3: Nagao et al. (2004).
\item[$^{\mathrm{f}}$]
  Observed line width is comparable to the instrumental broadening width.
\end{list}
\end{table*}

We have already obtained optical spectra of 3 
NB921-depressed $i^\prime$-dropout galaxies; 
SDF J132440.6+273607 (\#4; $z=6.33$; Nagao et al. 2004), 
SDF J132426.5.0+271600 and SDF J132442.5+272423 (\#3 
and \#5; $z=6.03$ and 6.04; Nagao et al. 2005a). Among 
the remaining 5 NB921-depressed $i^\prime$-dropout 
galaxies, we observed 4 objects on 26 April 2006 (UT). 
This spectroscopy was carried out with the Faint
Object Camera And Spectrograph (FOCAS; Kashikawa et al. 
2002) on the Subaru telescope (Iye et al. 2004), in 
its multi-object slit mode. 
The four objects were chosen to optimize the slit 
mask design (by also accomodationg several other objects 
observed within the broader context of the SDF project).
The 175 lines mm$^{-1}$ 
echelle grating and the SDSS $z^\prime$ filter were used. 
The resulting wavelength coverage was 
$\sim 8300-10000\mbox{\rm \AA}$, with a dispersion of 
$\sim 0.95\mbox{\rm \AA}$ pixel$^{-1}$. The adopted slit 
width was 0.83 arcsec, giving a spectral resolution of 
$R \sim 1500$ or $\Delta \lambda \sim 6 \mbox{\rm \AA}$ at 
$\sim 9000\mbox{\rm \AA}$ as measured from the widths of 
atmospheric OH emission lines. The spatial sampling was 
0.31 arcsec per resolution element, as we adopted 3 pixel 
on-chip binning. The seeing was variable during the 
observation (0.5 -- 1.0 arcsec). The total integration 
times were 7200 sec for SDF J132345.6+271701 (\#1) and 
SDF J132422.0+271742 (\#2), and 9000 sec for 
SDF J132519.4+271829 (\#6) and SDF J132526.1+271902
(\#8). 
Note that the position angle of the slits was set
to the North-South direction for all the target objects.
Since no neighboring objects are around the target
objects (Figure 2), the detected signal is believed
to be from the targets.
We also obtained spectra of the spectrophotometric 
standard star Feige 34 (Oke 1990) for flux calibration. 
The obtained data were reduced in the standard manner, 
by using IRAF.

\section{Results}

Among the four observed objects, SDF J132345.6+271701
(\#1) and SDF J132519.4+271829 (\#6) show a strong 
emission line in their spectra, 
whose peak wavelengths are at 8634$\mbox{\rm \AA}$ and 
8512$\mbox{\rm \AA}$, respectively.
In Figures 3 and 4, 
the sky-subtracted position-velocity spectrogram, the 
extracted one-dimensional spectrum, and the typical 
sky spectrum are shown for these two objects, 
respectively. The aperture size used for the extraction 
of the one-dimensional spectrum is 5 binned pixels 
($\sim$1.6 arcsec). No continuum emission is detected 
for either object. Both emission lines show a clear 
asymmetric profile, i.e., a sharp decline on the blue 
side and a prominent tail on the red side, which 
suggests that the observed emission lines are Ly$\alpha$.
Note that if the detected emission line were H$\beta$, 
[O {\sc iii}]$\lambda$5007 or H$\alpha$, other 
rest-frame optical emission line(s) should be seen in 
the observed wavelength range. If the detected emission 
line was [O {\sc ii}]$\lambda$3727, it should be 
resolved as a doublet emission, since the expected 
wavelength separation of the redshifted 
[O {\sc ii}]$\lambda\lambda$3726,3729 is 
$\sim 6 \mbox{\rm \AA}$ (or $\sim$6 pixels), which 
corresponds to $\sim 210$ km s$^{-1}$. Here we assume 
that the velocity width of the [O {\sc ii}] lines is
not very broad; otherwise we could not resolve the
[O {\sc ii}] doublet. However our assumption seems
valid for normal star-forming galaxies,
because the [O {\sc ii}] doublet of some star-forming
galaxies in the SDF is indeed resolved in our previous
spectroscopic follow-up observations (Shimasaku et al.
2006; Ly et al. 2007).
The fact that we see only one strong emission line 
with an asymmetric profile in each spectrum
strongly suggests
that we have detected Ly$\alpha$ at $z \sim 6$. 

To quantify the asymmetry of the detected emission 
lines, we calculated two independent parameters;
$f_{\rm red}/f_{\rm blue}$ and $S_w$. The former is 
the ratio between $f_{\rm red}$ and $f_{\rm blue}$, 
where $f_{\rm red}$ is the flux at wavelengths longer 
than the emission-line peak, while $f_{\rm blue}$ is 
that at shorter wavelengths (Taniguchi et al. 2005; 
Nagao et al. 2005a; see also Haiman 2002 for a 
theoretical discussion on the parameter 
$f_{\rm red}/f_{\rm blue}$). The measured ratios are 
2.08$\pm$0.12 and 2.47$\pm$0.30 for 
SDF J132345.6+271701 (\#1) and SDF J132519.4+271829
(\#6), respectively. Note that the latter value may 
be overestimated, since the OH airglow emission at 
the blue side of the Ly$\alpha$ emission may be
over-subtracted (Fig.4). The $f_{\rm red}/f_{\rm blue}$ 
ratio of both line-detected objects is significantly 
larger than unity, and also larger than that of most 
LAEs reported by Taniguchi et al. (2005). This result 
is consistent with the interpretation that the detected 
emission lines are Ly$\alpha$. The latter parameter 
($S_w$) is the weighted skewness (Shimasaku et al. 2006; 
Kashikawa et al. 2006), which is larger for objects 
with higher asymmetries and/or larger emission-line 
widths. The calculated values are 4.00$\pm$0.22 and 
8.72$\pm$0.28 for SDF J132345.6+271701 (\#1) and 
SDF J132519.4+271829 (\#6), respectively. 
Since the weighted skewness of emission lines 
arising from possible low-redshift interlopers 
generally do not exceed 3 (Kashikawa et al. 
2006; see also Shimasaku et al. 2006), 
the derived values of the weighted skewness are also 
consistent with the interpretation that the detected 
emission lines are Ly$\alpha$. Taking the derived 
values of $f_{\rm red}/f_{\rm blue}$ and $S_w$ for the 
two line-detected objects into account, we conclude 
that the detected emission lines are Ly$\alpha$ and 
accordingly that the redshifts of the two objects are 
6.11 and 6.00, respectively. The observed Ly$\alpha$ 
fluxes are (4.5$\pm$0.1)$\times$10$^{-17}$ ergs 
s$^{-1}$ cm$^{-2}$ and (3.4$\pm$0.2)$\times$10$^{-17}$ 
ergs s$^{-1}$ cm$^{-2}$, without any correction for
slit loss. The Ly$\alpha$ luminosities are thus 
calculated to be $1.9 \times 10^{43}$ ergs s$^{-1}$ 
and $1.3 \times 10^{43}$ ergs s$^{-1}$, respectively.
The measured emission-line widths in FWHM are 
$11.3^{+1.1}_{-1.6}\mbox{\rm \AA}$ and 
$8.6^{+1.9}_{-5.7}\mbox{\rm \AA}$. These values correspond to 
$9.6^{+1.3}_{-2.0}\mbox{\rm \AA}$ and 
$6.2^{+2.4}_{-6.2}\mbox{\rm \AA}$, or velocity widths of 
$330^{+50}_{-70}$ km s$^{-1}$ and 
$220^{+80}_{-220}$ km s$^{-1}$, respectively, after 
corrected for the instrumental broadening effect.
Here the absorption effects are 
not taken into account.

Adopting these redshifts, the N{\sc v}$\lambda$1240 
emission line would be expected at $8816\mbox{\rm \AA}$ 
and $8680\mbox{\rm \AA}$ for SDF J132345.6+271701 (\#1) 
and SDF J132519.4+271829 (\#6) respectively, if these 
two objects were active galactic nuclei (AGNs).
However, no emission-line features are seen at the
corresponding wavelengths (see Figures 3 and 4).
The 3$\sigma$ upper limits for the N {\sc v} emission 
are $5.7 \times 10^{-18}$ ergs s$^{-1}$ cm$^{-2}$ and
$6.9 \times 10^{-18}$ ergs s$^{-1}$ cm$^{-2}$ for 
these two objects respectively, assuming the same 
velocity widths as Ly$\alpha$. However, we cannot 
completely rule out the possibility that one or both 
of these two objects are AGNs, since some high-$z$ 
narrow-line radio galaxies show very weak N~{\sc v} 
compared with Ly$\alpha$ (e.g., De Breuck et al. 2000; 
Nagao et al. 2006; see also Malkan et al. 1996).

\begin{figure}
\centering
\includegraphics[width=8.3cm]{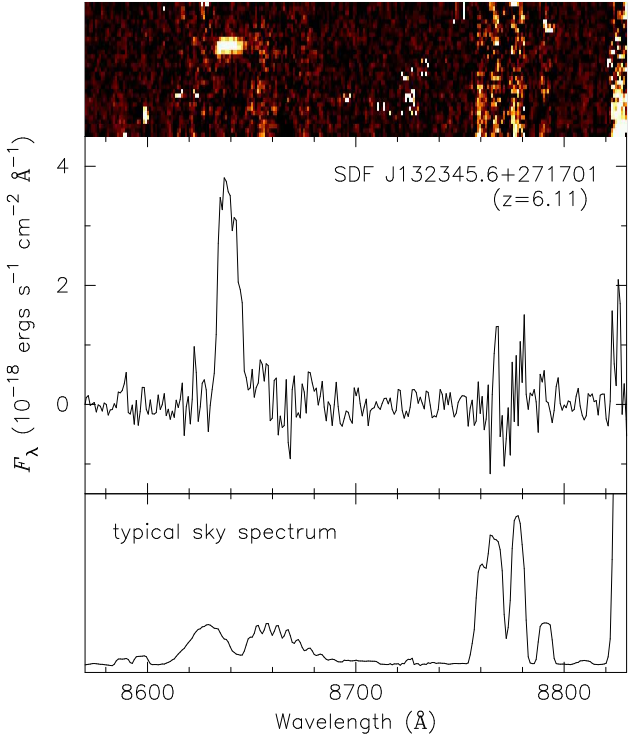}
\caption{
   Sky-subtracted optical position-velocity spectrogram ($top$) 
   and one-dimensional spectrum ($middle$) of SDF J132345.6+271701 
   (\#1) obtained with FOCAS on Subaru. The typical 
   spectrum of the sky emission is also shown in the bottom panel.
}
\label{fig03}
\end{figure}

\begin{figure}
\centering
\includegraphics[width=8.3cm]{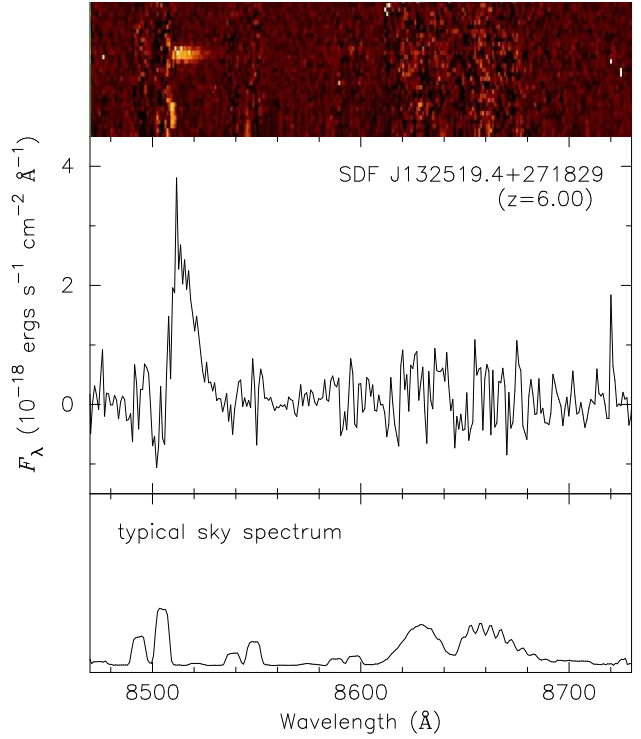}
\caption{
   Same as Figure 3 but for SDF J132519.4+271829 (\#6).
}
\label{fig04}
\end{figure}

As mentioned in \S1, it is sometimes very difficult 
to estimate the Ly$\alpha$ equivalent width because 
measuring the continuum in the low-S/N spectra is
very difficult. However, we can determine the flux 
density of the continuum emission for those two galaxies 
from the NB921-band magnitude, since the NB921-band flux 
does not contain Ly$\alpha$ flux, unlike the case of 
LAEs selected by the NB921 excess (e.g., 
Kodaira et al. 2003; Taniguchi et al. 2005). The 
NB921-band magnitudes of the two galaxies are 26.37 mag 
and 26.38 mag, which correspond to the flux densities of
$f_\lambda =$ $(4.1 \pm 1.1) \times 10^{-20}$ 
ergs s$^{-1}$ cm$^{-2}$ $\mbox{\rm \AA}^{-1}$ and 
$(4.2 \pm 1.2) \times 10^{-20}$ ergs s$^{-1}$ cm$^{-2}$ 
$\mbox{\rm \AA}^{-1}$ at the wavelengths of the Ly$\alpha$ 
peak, respectively. Then the rest-frame equivalent 
widths of the Ly$\alpha$ emission are calculated to be 
$153 \pm 42 \mbox{\rm \AA}$ and $114 \pm 32 \mbox{\rm \AA}$ for 
SDF J132345.6+271701 (\#1) and SDF J132519.4+271829 
(\#6) respectively, assuming a flat UV continuum. Note 
that if the detected emission lines were [O {\sc ii}], 
the rest-frame equivalent widths would be 
470$\mbox{\rm \AA}$ and 349$\mbox{\rm \AA}$, respectively. These 
large values are quite rare for [O {\sc ii}] emitters 
(e.g., Ajiki et al. 2006), which also supports our 
interpretation that the detected emission lines are 
Ly$\alpha$.

The obtained spectroscopic properties are summarized 
in Table 2. In the same table, properties of our 
previous spectra (Nagao et al. 2004, 2005a) are also 
given for the reader's convenience. Note that there 
are no spectral features in the obtained spectra of 
SDF J132422.0+271742 (\#2) and SDF J132526.1+271902
(\#8). We will discuss these non-detections briefly 
in the next section.
In this table, we also give the minimum $EW_0$(Ly$\alpha$) 
that can be detected by the NB921-depressed $i^\prime$-dropout 
method for galaxies with the redshift and the $z^\prime$-band 
magnitude of each NB921-depressed $i^\prime$-dropout galaxy. 
We will discuss this quantity in \S\S4.1.

\section{Discussion}

\subsection{The success rate of the NB921-depressed
            i-dropout selection method}

In this observing run we observed four NB921-depressed 
$i^\prime$-dropout galaxies, and found strong emission 
lines in two of them. Could the two remaining 
emission-line undetected objects also be strong 
LAEs at $6.0 < z < 6.5$, as expected by the 
NB921-depressed $i^\prime$-dropout selection?
Since high-$z$ Ly$\alpha$ emission shows a clear
asymmetric profile and especially a prominent tail
toward the red side of the emission-line peak,
the Ly$\alpha$ emission of LAEs is expected to be 
resolved with the current wavelength resolution 
$R \sim 1500$. This means that the Ly$\alpha$ emission 
should be found even when the redshifted Ly$\alpha$ 
line falls on an isolated (i.e., unresolved) OH airglow 
emission line. However, the Ly$\alpha$ detection would 
be difficult when the Ly$\alpha$ emission line falls on 
blended OH lines, such as the lines at 
$8760\mbox{\rm \AA} \la \lambda \la 8780\mbox{\rm \AA}$ seen at 
the bottom panel of Figure 3 (which corresponds to a 
redshift range $6.21 \la z \la 6.22$).
SDF J132422.0+271742 (\#2) has $z^\prime$ and $NB921$
magnitudes which are the faintest among the 
NB921-depressed $i^\prime$-dropout galaxy sample. Thus 
the non-detection of the emission line might be simply 
due to an insufficient integration time. On the 
contrary, the $z^\prime$ and $NB921$ magnitudes of 
SDF J132526.1+271902 (\#8) are the brightest among
the NB921-depressed $i^\prime$-dropouts. Since this 
object is so bright, it is selected as an 
NB921-depressed object in spite of its small amount of 
the NB921 depression, $z^\prime - NB921 = -0.25$. 
Therefore the expected Ly$\alpha$ equivalent width of 
this object is so small 
(which is estimated to be $\sim 50 \mbox{\rm \AA}$ in
the rest frame by assuming $z=6.0$ and a flat UV slope)
that the detection of 
Ly$\alpha$ could be more difficult than the other 
NB921-depressed $i^\prime$-dropout galaxies. Taking 
all of these considerations into account, we cannot 
reject the possibility that the two Ly$\alpha$ 
non-detected objects are also LAEs at $6.0 < z < 6.5$. 
Deeper spectroscopic observations are necessary to 
investigate these objects further.

Therefore, the ``success rate'' of the NB921-depressed
$i^\prime$-dropout selection method (i.e., the 
probability that this selection identify LAEs at 
$6.0 < z < 6.5$ correctly) is between 5/8 (63\%) and 
8/8 (100\%). The rest-frame Ly$\alpha$ equivalent width 
of all of the five Ly$\alpha$-detected objects is quite 
large, at least $\sim 100 \mbox{\rm \AA}$, as also expected 
by the selection method (see Table 2). It should be 
mentioned here that, as discussed in Nagao et al. 
(2005a), the NB921 depression is not expected in 
Galactic late-type stars and thus this selection method 
is not contaminated by stars, unlike a simple 
$i^\prime$-dropout sample (see, e.g., Stanway et al. 
2004). 

\begin{figure}
\centering
\rotatebox{-90}{\includegraphics[width=4.8cm]{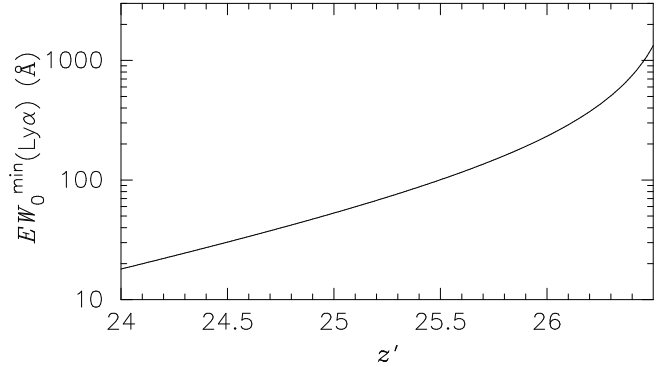}}
\caption{
   Minimum $EW_0$(Ly$\alpha$) that can be selected by 
   the NB921-depressed $i^\prime$-dropout method 
   [$EW_0^{\rm min}$(Ly$\alpha$)] as a function of the
   $z^\prime$-band magnitude, estimated by assuming 
   $z = 6.0$ and a flat UV slope.
}
\label{fig05}
\end{figure}

Here we discuss on the limiting $EW$(Ly$\alpha$) of the 
NB921-depressed $i^\prime$-dropout selection method. This 
selection technique does not select only LAEs with a large 
$EW_0$(Ly$\alpha$) in principle, because the selection 
criterion on the NB921 depression is based on the sky
error as described in \S2. This is different from usual LAE
surveys based on the narrow-band excession, for which a 
certain excess magnitude is generally used instead of the 
sky noise. Therefore the NB921-depressed $i^\prime$-dropout 
method would select LAEs with a relatively small 
$EW_0$(Ly$\alpha$) if the imaging data were enough deep and 
accordingly the sky noise was very small. In other words, 
the minimum $EW_0$(Ly$\alpha$) that can be selected by the
NB921-depressed $i^\prime$-dropout method [hereafter 
$EW_0^{\rm min}$(Ly$\alpha$)] depends on the magnitude of 
target galaxies, since the $z^\prime - NB921$ color of 
brighter objects is less affected by the sky error with
respect to that of fainter objects. To see this effect
quantitatively, we show the dependence of 
$EW_0^{\rm min}$(Ly$\alpha$) on the $z^\prime$ magnitude 
for the case of the SDF dataset in Figure 5, where
$z=6.0$ and a flat UV slope are assumed. It is demonstrated 
that only LAEs with $EW_0$(Ly$\alpha$) $\ga$ 100$\mbox{\rm \AA}$ 
are selected among faint galaxies with $z^\prime \ga 25.5$, 
while LAEs with $EW_0$(Ly$\alpha$) $\sim$ a tens of 
10$\mbox{\rm \AA}$ could be also selected among relatively 
bright galaxies with $z^\prime \la 25.0$. Note that the 
latter case corresponds to the case of SDF J132526.1+271902 
(\#8). To compare $EW_0^{\rm min}$(Ly$\alpha$) and the
detected $EW_0$(Ly$\alpha$) for each NB921-depressed 
$i^\prime$-dropout galaxy, we also give 
$EW_0^{\rm min}$(Ly$\alpha$) for each set of the redshift 
and the $z^\prime$-band magnitude of 5 spectroscopically 
identified NB921-depressed $i^\prime$-dropout galaxies in 
Table 2. In order to carry out systematic surveys for 
strong LAEs utilizing the NB921-depressed $i^\prime$-dropout
method, it would be more appropriate to adopt a certain 
$z^\prime - NB921$ depression magnitude as a selection 
criterion rather than using the sky error. The reason 
for adopting
the selection criterion based on the sky noise 
is that one of our motivations in the photometric selection 
is to construct a sample of the target objects for our
spectroscopic observations to find candidates of 
PopIII-hosting galaxies (see Nagao et al. 2005b). Due to 
the limited observing time, we had to focus on convincing 
candidates, i.e., objects with a significant NB921 
depression with respect to the sky error.

\subsection{Implication for the stellar population}

The rest-frame Ly$\alpha$ equivalent width of the 
NB921-depressed $i^\prime$-dropout galaxies with a 
Ly$\alpha$ detection (given in Table 2) ranges from 
94$\mbox{\rm \AA}$ to 236$\mbox{\rm \AA}$. Although these 
values are less than the ``critical value'' 
$EW_0$(Ly$\alpha$) = 240$\mbox{\rm \AA}$ above which it 
cannot be explained by normal stellar populations
(e.g., Malhotra \& Rhoads 2002), it should be noted 
that the values we obtained are not corrected for 
absorption effects. 
The amount of the Ly$\alpha$ absorption depends on 
various parameters including the neutral fraction of the 
inter-galactic matter (IGM) and the kinematic status of 
neutral hydrogen within LAEs themselves (e.g., Neufeld 1990; 
Haiman 2002; Mas-Hesse et al. 2003; Ahn 2004). Here it 
should be kept in mind that the IGM is not perfectly 
re-ionized at $z \sim 6$, which is suggested by recent
observations of high-$z$ quasars (Fan et al. 2006) and LAEs
(Kashikawa et al. 2006). It has been extensively argued 
whether Ly$\alpha$ photons from galaxies at an earlier 
epoch than the completion of the cosmic re-ionization are 
observable or not. Some calculations suggest that the 
Ly$\alpha$ photons from such galaxies are strongly 
suppressed (e.g., Miralda-Escude 1998; Miralda-Escude \&
Rees 1998; Loeb \& Rybicki 1999) and other theoretical 
works infer that a large fraction of the Ly$\alpha$ 
photons can transmit due to cosmological H {\sc ii} 
regions (e.g., Madau \& Rees 2000; Cen \& Haiman 2000). 
Haiman (2002) investigated various parameter dependences 
of the Ly$\alpha$ transmission fraction for high-$z$ LAEs 
by taking both the neutral hydrogen within LAEs themselves 
and the neutral IGM into account. They showed that the 
ratio of the transmitted Ly$\alpha$ flux to the total 
(intrinsic) Ly$\alpha$ flux 
[$F_0$(Ly$\alpha$)/$F_{\rm tot}$(Ly$\alpha$)] is estimated 
to be $\sim$10\% for typical LAEs with a star-formation 
rate (SFR) $\sim$ 10 $M_\odot$ yr$^{-1}$ and $\sim$50\% 
for LAEs with $SFR \sim 100 M_\odot$ yr$^{-1}$. 
Following this result, we assume 
$F_0$(Ly$\alpha$)/$F_{\rm tot}$(Ly$\alpha$) $\sim$0.5 to 
correct the absorption effect on the observed Ly$\alpha$
flux rather conservatively, in the sense that the actual
$F_0$(Ly$\alpha$)/$F_{\rm tot}$(Ly$\alpha$) of our 
spectroscopic sample would be smaller than 0.5 since the 
SFR is less than 100 $M_\odot$ yr$^{-1}$ 
(Nagao et al. 2004, 2005a). 
See also, e.g., Santos (2002) and Dijkstra et al. (2006)
for the justification of our adopted value of 
$F_0$(Ly$\alpha$)/$F_{\rm tot}$(Ly$\alpha$).
Interestingly, by adopting 
this assumption, the intrinsic $EW_0$(Ly$\alpha$) for 
three out of the five spectroscopically identified 
NB921-depressed $i^\prime$-dropout galaxies is then 
expected to exceed the critical value $EW_0$(Ly$\alpha$) 
= 240$\mbox{\rm \AA}$. This result is consistent with the idea 
that galaxies at $z \ga 6$ contains young stellar 
populations, whose age may be younger than 
$\sim 10^7$ years as discussed by Malhotra \& Rhoads 
(2002); see also Shimasaku et al. (2006).

The huge intrinsic Ly$\alpha$ equivalent width would
indicate that the NB921-depressed $i^\prime$-dropout 
galaxies may contain a significant number of PopIII
stars. Therefore our NB921-depressed 
$i^\prime$-dropout galaxy sample offers fascinating 
targets for observational searches for PopIII 
stars, which will be an 
important topic in the forthcoming decade (see, e.g., 
Jimenez \& Haiman 2006). We have already searched 
for an observational signature in one of the 
NB921-depressed $i^\prime$-dropout galaxies,
SDF J132440.6+273607 (\#4; Nagao et al. 2005b), 
through He {\sc ii}$\lambda$1640 emission (see, e.g.,
Tumlinson \& Shull 2000; Oh et al. 2001; 
Schaerer 2002, 2003). Although the 
He {\sc ii}$\lambda$1640 emission was not detected
in this object, similar observations will provide
important constraints on theoretical PopIII models
(e.g., Scannapieco et al. 2003; Nagao et al. 2005b).

\subsection{Evolution of the Ly alpha equivalent width
            distribution}

It has been known that narrow-band selected galaxies 
(e.g., emission-line galaxies) at high redshift tend 
to have high Ly$\alpha$ equivalent widths, which 
sometimes exceeds the critical value of
$EW_0$(Ly$\alpha$) = 240$\mbox{\rm \AA}$ (e.g., 
Malhotra \& Rhoads 2002; Shimasaku et al. 2006).
However, some broad-band selected high-$z$ galaxies 
such as Lyman-break galaxies (LBGs) 
rarely
have high Ly$\alpha$ equivalent widths. For instance, 
at $z \sim 3$ the fraction of LBGs showing 
$EW_0$(Ly$\alpha$) $>$ 100$\mbox{\rm \AA}$ is $\approx$1\% 
of $\sim$1000 galaxies (Shapley et al. 2003). Since 
our sample is also broad-band selected, it is 
interesting to compare the fraction of galaxies with 
a large equivalent width to investigate whether the 
stellar population of galaxies evolves as a function 
of redshift. At $z \sim 6$, our $i^\prime$-dropout 
galaxy sample (48 objects) contains at least 5 
galaxies with $EW_0$(Ly$\alpha$) $\ga$ 100$\mbox{\rm \AA}$ 
[here we retain SDF J132426.5+271600 (\#3) as a 
large EW object although its EW is slightly below
the criterion]. However, $z^\prime$-band flux of the
NB921-depressed $i^\prime$-dropout galaxies is 
strongly boosted by the Ly$\alpha$ contamination.
After correction for this effect, they may fail to 
satisfy the $i^\prime$-dropout criterion, 
$i^\prime - z^\prime > 1.5$. Indeed if adopting the 
NB921 magnitude instead of $z^\prime$ as a continuum 
magnitude at the long side of the Lyman break,
SDF J132442.5+272423 (\#5) should be removed from 
the SDF $i^\prime$-dropout galaxy sample 
($i^\prime - NB921 = 0.98$). Therefore we estimate 
the fraction of $i^\prime$-dropout galaxies having 
$EW_0$(Ly$\alpha$) $\ga$ 100$\mbox{\rm \AA}$ is 4/48 
$\approx$ 8\%. Note that this estimated fraction is 
a conservative lower limit, because (1) some 
Ly$\alpha$ non-detected NB921-depressed 
$i^\prime$-dropout galaxies may also have such a high 
equivalent width as discussed above, and (2) some of 
the 48 $i^\prime$-dropout objects may be 
emission-line galaxies at lower redshift or Galactic 
late-type stars, since most of the SDF 
$i^\prime$-dropout objects have not been confirmed by 
spectroscopic follow-up observations (see, e.g., 
Stanway et al. 2004).

The derived lower limit of the fraction of 
$i^\prime$-dropout objects having $EW_0$(Ly$\alpha$) 
$\ga$ 100$\mbox{\rm \AA}$ is significantly higher than 
the value at $z \sim 3$. However, it should be noted 
that the limiting magnitude of these two samples are 
different; that is, while the spectroscopic survey of
LBGs at $z \sim 3$ by Shapley et al. (2003) reaches 
down to $R_{\rm AB} \sim 25.5$, our SDF 
$i^\prime$-dropout sample consists of galaxies with 
$z^\prime < 26.1$. These limiting magnitudes 
correspond to the absolute UV magnitudes of 
$M_{1500} =$ --20.0 and --20.6 respectively, and thus 
the galaxies in the $z \sim 6$ sample are 
intrinsically brighter with respect to those in the 
$z \sim 3$ sample, systematically. Recently, 
Ando et al. (2006) reported that the LBGs with a high 
Ly$\alpha$ equivalent width are rarer in brighter 
samples than in fainter samples, at $z \ga 5$.
Shapley et al. (2003) also reported based on 
their huge number of a LBG spectroscopic sample that 
the broad-band magnitude of LBGs with a stronger 
Ly$\alpha$ emission tends to be fainter than that of
LBGs with a weaker Ly$\alpha$ emission (or with a 
Ly$\alpha$ absorption instead of emission; see 
Table 3 of Shapley et al. 2003).
Therefore, taking the difference in the limiting 
magnitude between the samples at $z \sim 6$ and at 
$z \sim 3$ and the dependence of the equivalent width 
on the luminosity into account, the difference in the 
fraction of galaxies with $EW_0$(Ly$\alpha$) $\ga$ 
100$\mbox{\rm \AA}$ between at $z \sim 6$ and at 
$z \sim 3$ should be even more significant. 
Note that the significance may be intrinsically
much more if the IGM effect against the Ly$\alpha$ 
transmission is stronger at $z\sim6$ than at $z\sim3$.
This result is consistent with a recent study by 
Shimasaku et al. (2006) that the fraction of LBGs 
having $EW_0$(Ly$\alpha$) $\ga$ 100$\mbox{\rm \AA}$ 
significantly evolves from $z \sim 3$ to $z \sim 6$.
All of these results are consistent with the idea 
that the typical stellar population of galaxies is 
significantly younger at $z \sim 6$ than that at 
$z \sim 3$.
This may also be consistent with recent
findings that the slope of the rest-frame UV 
continuum of some galaxies is bluer at $z \sim 6$ 
than the typical UV slope at $z \sim 3$ (e.g., 
Stanway et al. 2004, 2005; Bouwens et al. 2005; 
Yan et al. 2005).

\section{Summary}

In earlier papers we identified 8 NB921-depressed 
$i^\prime$-dropout galaxies, which are expected to be
strong LAEs at $6.0 \la z \la 6.5$, through the 
narrow-band and broad-band photometric data of the SDF
(Nagao et al. 2004, 2005a). In addition to 3 
previously spectroscopically confirmed ones,
we found that other two NB921-depressed $i^\prime$-dropout 
galaxies are also LAEs with $EW_0$(Ly$\alpha$) $>$ 
100$\mbox{\rm \AA}$ at $z = 6.11$ and 6.00, by new optical
spectroscopy.
This result combined with our previous spectroscopic runs
means that at least 5 objects among 8 NB921-depressed 
$i^\prime$-dropout galaxies are indeed LAEs 
having $EW_0$(Ly$\alpha$) $\ga$ 100$\mbox{\rm \AA}$ 
at $6.0 \la z \la 6.5$; these results suggest that the
NB921-depressed $i^\prime$-dropout selection method is
an efficient technique to identify strong LAEs in a
wide redshift range, $6.0 \la z \la 6.5$.

The obtained results also suggest that more than 8\% of
the $i^\prime$-dropout galaxies in the SDF have
a large Ly$\alpha$ equivalent width of $EW_0$(Ly$\alpha$) 
$\ga$ 100$\mbox{\rm \AA}$. This is in contrast with LBGs at
$z \sim 3$, where such strong LAEs are much rarer 
($\sim$1\%). This also implies a strong redshift evolution
in the Ly$\alpha$ equivalent width distribution
from $z \sim 6$ to $z \sim 3$, consistently with a
stellar population of broad-band selected galaxies 
which is significantly younger at $z \sim 6$ than
at $z \sim 3$.

\begin{acknowledgements}
  This research is based on data collected at the Subaru 
  telescope, which is operated by the National Astronomical 
  Observatory of Japan. We are grateful to the staff of the 
  Subaru telescope and to the all members of the Subaru Deep 
  Field project. We also thank the anonymous referee
  for useful comments. TN and SSS are financially supported 
  by the Japan Society for the Promotion of Science (JSPS) 
  through JSPS Research Fellowship for Young Scientists.
\end{acknowledgements}

\end{document}